\documentclass[sigconf]{acmart}

\usepackage{xcolor}
\usepackage[most]{tcolorbox}
\usepackage[table]{xcolor}
\definecolor{lightgreen}{RGB}{198, 226, 198}
\usepackage{booktabs}
\usepackage{multirow}
\usepackage{todonotes}
\usepackage{makecell}
\definecolor{promptdarkgreen}{HTML}{85A92D}
\definecolor{lightgreen}{HTML}{F3F5EE}

\newif\ifshowcomments
\showcommentstrue  % Comment this line to turn off comments 

\usepackage{stfloats} % <- Fix figure* floating bug that flushes it to next page

\setcopyright{none}
\renewcommand\footnotetextcopyrightpermission[1]{}

\AtBeginDocument{%
  \providecommand\BibTeX{{%
    \normalfont B\kern-0.5em{\scshape i\kern-0.25em b}\kern-0.8em\TeX}}}

\graphicspath{{./images/}}

\begin{document}

%%
%% The "title" command has an optional parameter,
%% allowing the author to define a "short title" to be used in page headers.
\title{Towards Behavior Tree–Guided\\ Vulnerability Detection with Lightweight LLMs}

%%
%% The "author" command and its associated commands are used to define
%% the authors and their affiliations.
%% Of note is the shared affiliation of the first two authors, and the
%% "authornote" and "authornotemark" commands
%% used to denote shared contribution to the research.

\author{Enna Basic}
\affiliation{%
  \institution{Department of Computer Science}
  \city{Örebro University}
  \country{Sweden}
 }
  \email{enna.basic@oru.se}

\author{Alberto Giaretta}
\affiliation{%
  \institution{Department of Computer Science}
  \city{Örebro University}
  \country{Sweden}
 }
  \email{alberto.giaretta@oru.se}

%%
%% By default, the full list of authors will be used in the page
%% headers. Often, this list is too long, and will overlap
%% other information printed in the page headers. This command allows
%% the author to define a more concise list
%% of authors' names for this purpose.
\renewcommand{\shortauthors}{Basic and Giaretta}

%%
%% The abstract is a short summary of the work to be presented in the
%% article.
\begin{abstract}

Large Language Models (LLMs) are increasingly used for software vulnerability detection, but their performance depends on how source code is represented in the input. Most prompting approaches use source code in its original form, while some works propose the use of structured representations. Abstract Syntax Trees (ASTs) are one of the most popular approaches, but AST verbosity increases input size relative to source code, making them hard to fit within some LLMs context windows. This paper investigates Behavior Trees (BTs) as an alternative intermediate representation for LLM-based vulnerability detection. BTs encode control flow, conditions, and executable actions more compactly than ASTs, making them a natural candidate when token count is a constraint. First, we propose a preprocessing stage that parses Java source code into ASTs and then converts them into BT representations.
We then compare vulnerability detection performance across 460 Java samples from the Juliet Java test suite, using three input representations: raw source code, AST, and BT. All experiments use a single quantized local LLM, Mistral Small 3.2 24B (Q4\_K\_M). Our results show that using BT representations improves recall on short code samples, while raw source code achieves higher precision. On longer samples, BTs improve overall performance over the original representation and fit within the context window, whereas many ASTs exceed the context limit. These findings suggest that BTs can provide a compact and useful structured representation for vulnerability detection with quantized, locally deployable LLMs.

\end{abstract}

%%
%% The code below is generated by the tool at: http://dl.acm.org/ccs.cfm
%% Please copy and paste the code instead of the example below.
%%


%%
%% Keywords. The author(s) should pick words that accurately describe
%% the work being presented. Separate the keywords with commas.

%% The following are not a requirement, delete if not using
% \received{20 February 2024}  %% inital submission date
% \received[revised]{12 March 2024} %% interim new draft
% \received[accepted]{5 June 2024}  %% publication version

%%
%% This command processes the author and affiliation and title
%% information and builds the first part of the formatted document.

\pagestyle{plain}

\maketitle

 \section{Introduction}

Large Language Models (LLMs) have been increasingly utilized for various software security tasks, including vulnerability detection. In vulnerability detection, an LLM can be prompted with a code sample and tasked to determine whether the code contains a security vulnerability and, when possible, identify the corresponding Common Weakness Enumeration (CWE) category~\cite{khare2025understanding,zhou2024large}.

Despite this potential, LLM-based vulnerability detection remains challenging. Prior work has shown that LLMs can fail to detect existing vulnerabilities, report vulnerabilities that are not present, or struggle with complex scenarios ~\cite{ullah2024llms, purba2023software, xia2025beyond}.

One factor that can influence LLM-based vulnerability detection is the way source code is represented in the input. Most prompting approaches provide source code in its original form, while other studies have explored structured representations such as Abstract Syntax Trees (ASTs), Code Property Graphs (CPGs), or enriched AST-based formats~\cite{wen2024scale,anbiya2025java,zhang2026vultrlm}. These representations can provide additional structural information, but they may also increase the input size and become difficult to use with limited context windows, especially for lightweight or locally deployed LLMs~\cite{sheng2025llms,li2025longcodeu}.

In this work, we investigate Behavior Trees (BTs) as an alternative intermediate representation for LLM-based vulnerability detection. BTs are commonly used to model behavior through hierarchical structures composed of control nodes, conditions, and actions~\cite{colledanchise2018behavior,iovino2022survey}. Since source code also contains execution logic, branching behavior, conditions, and executable operations, BTs offer a compact way to reorganize source code around behavior-oriented structure. We transform Java source code into BT representations and evaluate whether this representation can support vulnerability detection with a lightweight LLM.
To the best of our knowledge, BTs have not yet been studied as an input representation for LLM-based vulnerability detection.

The main contributions of this work are as follows:
\begin{itemize}

    \item We propose a behavior-tree-based preprocessing framework that transforms source code into behavior tree representations before feeding it to a lightweight LLM for vulnerability detection.
    
    \item We introduce a custom AST-to-BT conversion approach that maps Java source-code constructs into behavior tree nodes.

    \item We present a preliminary investigation of behavior tree representations as structured intermediate inputs for lightweight language models in vulnerability detection.

\end{itemize}

The remainder of this paper is organized as follows. Section~\ref{sec:background} presents background and related work on LLM-based vulnerability detection and structured code representations. Section~\ref{sec:method} describes the proposed BT-based preprocessing framework. Section~\ref{sec:experimental-setup} presents the experimental setup, including the datasets, evaluation strategy, and implementation details. Section~\ref{sec:results} reports the experimental results. Section~\ref{sec:discussion} discusses the main findings and their implications. Section~\ref{sec:threats} discusses threats to validity. Finally, Section~\ref{sec:conclusion} concludes the paper and outlines directions for future work.

\section{Background and Related Work}
\label{sec:background}

In this section, we provide background on the main concepts used in this work and discuss related studies on LLM-based vulnerability detection and structured code representations. First, we introduce LLM-based vulnerability detection as the broader application area. We then discuss ASTs as a commonly used structured representation of source code, followed by BTs as the representation investigated in this work.

\subsection{LLM-based Vulnerability Detection}

Security code review is a common practice for detecting vulnerabilities during software development; however, it can require significant time and effort in large-scale projects. This challenge becomes more difficult when many code changes are submitted continuously, and vulnerabilities may remain undetected~\cite{paul2021security}. Therefore, automated techniques that support reviewers in identifying potential security vulnerabilities can be valuable.

With the rapid development of LLMs, recent studies have explored their application to software engineering tasks, including vulnerability detection~\cite{liu2023software,zhou2024large,khare2025understanding}. In this context, LLMs are typically used to analyze source code and generate predictions about whether the code contains a security vulnerability.

Recent survey work further highlights that LLMs are rapidly being researched for software security tasks, including vulnerability detection, vulnerability repair, and security analysis~\cite{basic2024vulnerabilities,sheng2025llms}. However, these surveys also indicate that LLM-based vulnerability detection remains challenging, as models can still miss vulnerabilities, produce incorrect predictions, or struggle to provide reliable results across different settings.

Several studies have investigated the performance of LLMs in detecting vulnerabilities across various experimental settings. For example, prior work has evaluated the effect of prompt design, model choice, and vulnerability category on detection performance~\cite{zhou2024large,khare2025understanding}. Other studies have analyzed failure cases and reliability issues, showing that LLMs may miss vulnerable code, produce unstable predictions, or provide explanations that do not fully match the actual weakness~\cite{ullah2024llms,purba2023software,xia2025beyond}. Together, these studies suggest that LLM-based vulnerability detection is promising, but still sensitive to how the task is formulated and how the code is presented to the model.

One such design choice is how source code is represented in the model input. Prior work suggests that LLM-based vulnerability detection may be influenced not only by the model being used, but also by the representation provided to the model~\cite{wen2024scale,anbiya2025java}. This is important for the present work because different representations can expose different aspects of the code, such as syntax, control flow, or behavior-relevant structure.

% However, prior work suggests that LLM-based vulnerability detection may be influenced not only by the model being used, but also by how source code is represented in the input~\cite{wen2024scale,anbiya2025java}.

\subsection{Abstract Syntax Trees}

An Abstract Syntax Tree (AST) is a tree-based representation of the syntactic structure of source code, where nodes correspond to constructs in the program. ASTs are commonly produced during parsing and are widely used as intermediate representations in compilers and program analysis~\cite{zhang2019novel}. Since vulnerability detection can require reasoning about code structure and execution order, AST-based representations have been explored as a way to provide models with structural information from source code.

For instance, Anbiya et al.~\cite{anbiya2025java} investigated the use of structured code representations for Java vulnerability detection with LLMs, including AST, Code Property Graph (CPG), and combined AST-CPG representations. Their study demonstrates that vulnerability detection can be studied not only through the choice of model, but also through the choice of source-code representation.

Similarly, Wen et al.~\cite{wen2024scale} proposed SCALE, which builds an AST and augments nodes with short auto-generated comments. This representation combines code structure with natural-language descriptions, aiming to make vulnerability-relevant information more explicit in the model input. Zhang et al.~\cite{zhang2026vultrlm} further explored this direction through VulTrLM, which decomposes ASTs into smaller subtrees and enriches them with comments for LLM-assisted vulnerability detection. This also reflects the idea that structured representations can be used to reorganize source code before it is analyzed by an LLM.

% However, ASTs mainly represent the syntactic structure of source code, while this work investigates behavior trees as an alternative representation focused on program behavior and execution flow.

\subsection{Behavior Trees} 
\begin{figure}[tb]
\centering
\includegraphics[width=1\linewidth]{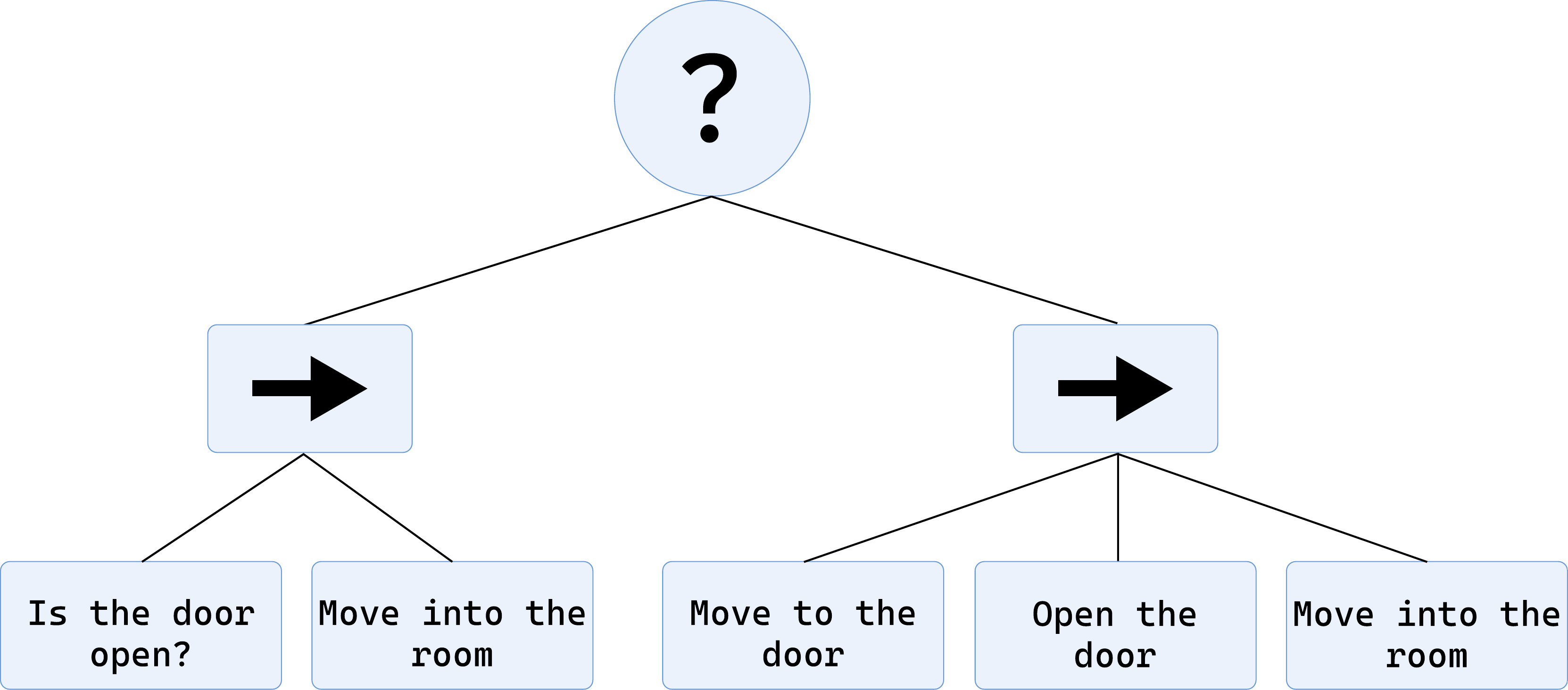}
\caption{Example of a simple BT illustrating Selector, Sequence, Condition, and Action nodes. Adapted from~\cite{sekhavat2017behavior}.}
\label{fig:bt_example}
\end{figure}

A Behavior Tree (BT) is a tree-based representation used to model execution logic and decision-making behavior. In a BT, internal nodes define how execution proceeds through the tree, while leaf nodes represent conditions or actions. 
Common BT node types include Sequence, Selector, Condition, and Action nodes. A Sequence node executes its children in order and succeeds only when all children succeed, while a Selector node represents alternative branches and succeeds when one of its children succeeds. Condition nodes evaluate whether a required condition holds, and Action nodes represent executable behavior. Figure~\ref{fig:bt_example} shows a simple BT example that illustrates these node types and how they are arranged in a tree structure.

BTs have been widely used in robotics, game AI, and autonomous systems, where complex behavior can be represented through reusable and hierarchical components~\cite{colledanchise2018behavior,iovino2022survey}.

While BTs have mainly been studied in robotics, game AI, and autonomous systems, their structure is relevant to this work because source code also contains execution logic, conditions, actions, and alternative branches. This makes BTs a potentially suitable representation for reorganizing source code around behavior-oriented structure rather than only syntactic structure. To the best of our knowledge, BTs have not yet been explored as an input representation for LLM-based vulnerability detection. This work, therefore, investigates whether BT-based representations of source code can serve as a useful structured input for this task.

\section{Method}
\label{sec:method}

\begin{figure*}[tb]
    \centering
    \includegraphics[width=0.9\textwidth]{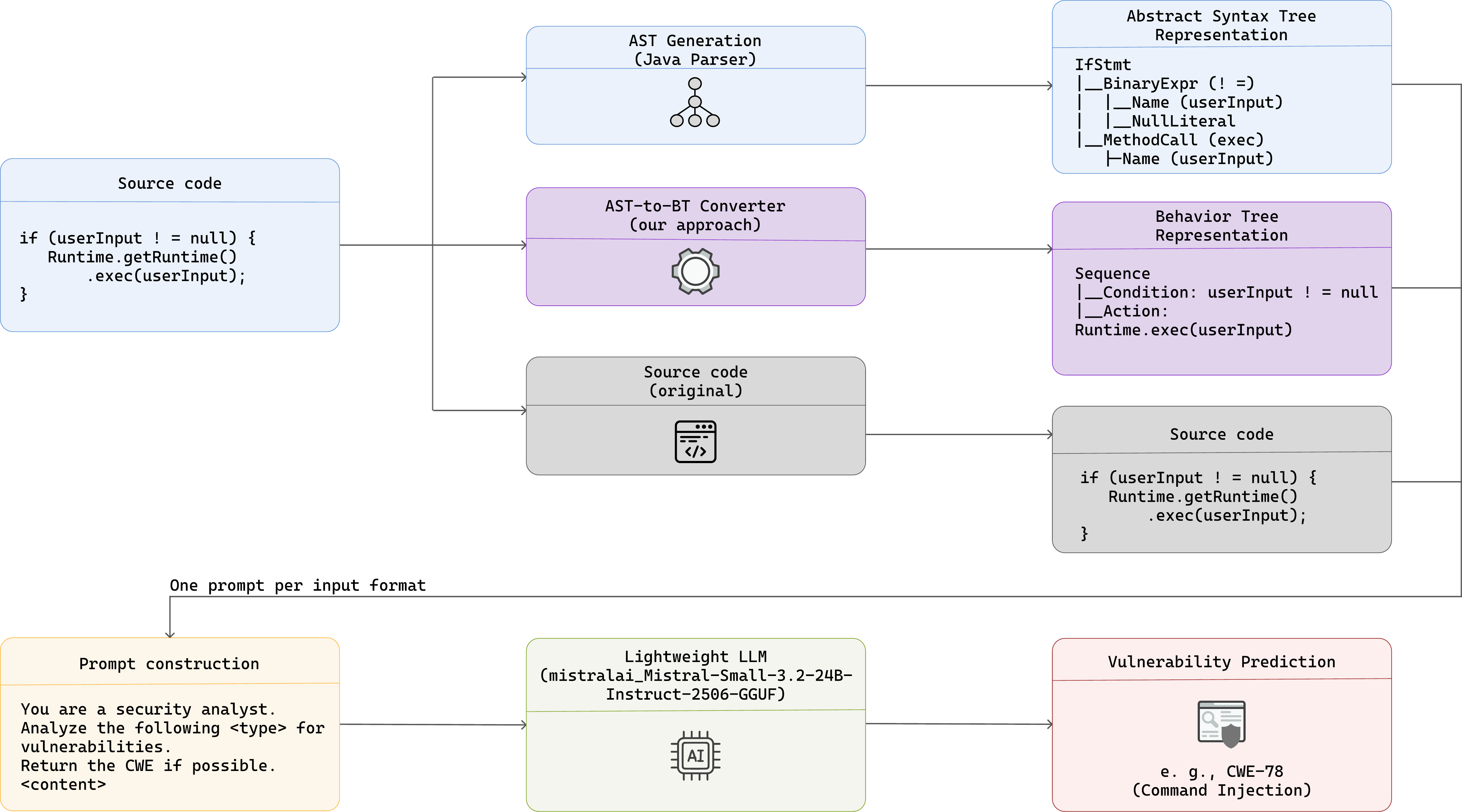}  
    \captionsetup{justification=centering}
    \caption{Overview of the proposed preprocessing framework.}
    \label{fig:framework-overview}
\end{figure*}

In this work, we propose a BT-based preprocessing framework for LLM-based vulnerability detection. Figure~\ref{fig:framework-overview} provides an overview of the proposed framework. The upper part of the figure shows how Java source code is transformed into the three input formats used in the experiments: source code in its original form, AST, and BT. The lower part shows how each input format is used with the shared prompt template and passed to the lightweight LLM for vulnerability prediction.

The framework first parses Java source code into an AST and then converts the AST into a BT representation using a custom converter. The original source code, the generated AST representation, and the generated BT representation are then used as three input formats in the experimental pipeline. For each of these input formats, a corresponding prompt is constructed using the same prompt template. The pipeline then passes the resulting prompt to a lightweight LLM. The model generates a vulnerability prediction, indicating whether the input is likely to contain a vulnerability and, when possible, identifying the corresponding CWE category.

The code snippet, prompt, and model name in Figure~\ref{fig:framework-overview} are included to illustrate the experimental setup used in this study, but the framework is not tied to this particular combination of code example, prompt wording, or model.

The converter maps relevant program constructs from the AST into BT nodes. Since BTs have not previously been explored as an input representation for LLM-based vulnerability detection, this converter is introduced to generate the BT representation used in this study.
The conversion is implemented through mapping rules that translate selected AST constructs into corresponding BT nodes, allowing the generated representation to capture the main control-flow organization of the source code. Table~\ref{tab:ast-bt-mapping} summarizes the main categories of mappings used by the converter.

\begin{table}[t]
\centering
\caption{Main AST-to-BT mappings used by the converter.}
\label{tab:ast-bt-mapping}

\begin{tabular}{p{0.42\linewidth}p{0.46\linewidth}}
\hline
\rowcolor{gray!15}
\textbf{AST construct} & \textbf{BT representation} \\
\hline
Class or method declaration & Sequence \\

\rowcolor{gray!15}
Sequential block & Sequence of child nodes \\

Conditional statement &
\makecell[l]{Selector with branch\\sequences} \\

\rowcolor{gray!15}
Boolean condition & Condition \\

Executable statement & Action \\

\rowcolor{gray!15}
Loop statement & Sequence with loop body \\
\hline
\end{tabular}
\end{table}

\begin{figure*}[tb]
    \centering
    \includegraphics[width=0.9\textwidth]{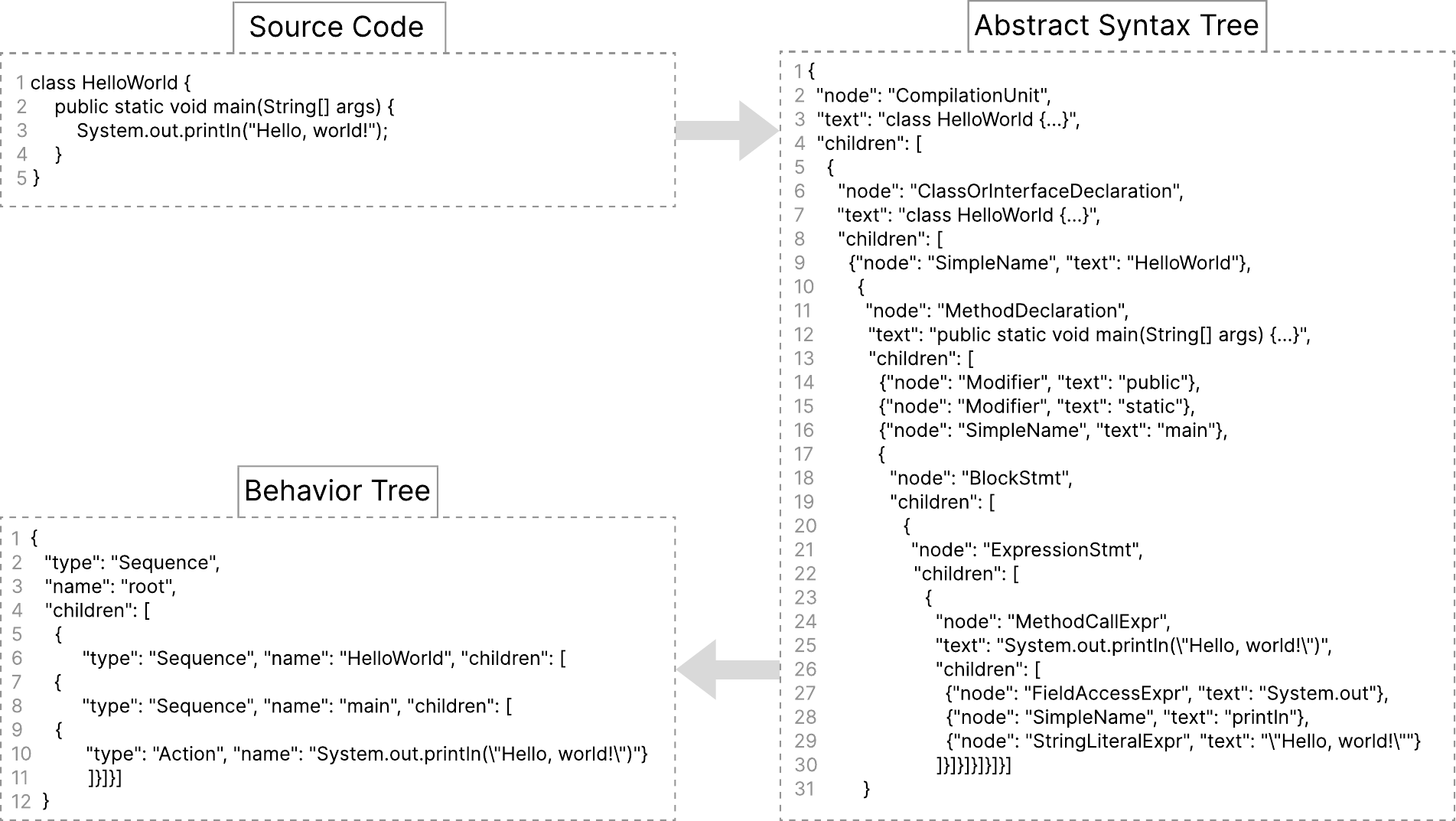}  
    \captionsetup{justification=centering}
    \caption{Illustrative example of a Java source code sample and its corresponding AST and BT representations in JSON format.}
    \label{fig:representation-example-json}
\end{figure*}

Figure~\ref{fig:representation-example-json} presents an illustrative Java example together with its generated AST and BT representations in JSON format. The example is included only to show the transformation process, but it also demonstrates how the AST representation can be more verbose, while the BT representation provides a more compact view focused on program structure and executable behavior.

\section{Experimental Setup}
\label{sec:experimental-setup}

This section describes the experimental design used to evaluate the proposed BT-based representation. We describe the datasets, evaluation strategy, and implementation details.

\subsection{Dataset}

The data used in this study is derived from the Juliet Java test suite from the NIST Software Assurance Reference Dataset.\footnote{\url{https://samate.nist.gov/SARD/test-suites/111}} Juliet contains synthetic Java vulnerability test cases organized by CWE categories. In this work, we use Java samples only and construct two smaller balanced datasets from selected CWE categories to evaluate vulnerability detection performance on code samples of different sizes. The datasets are balanced by CWE category and vulnerability label, with an equal number of vulnerable and non-vulnerable samples for each selected CWE.

The first dataset is the short-sample dataset, which we use to compare all three input representations: source code in its original form, AST, and BT. This dataset contains smaller code samples for which all three input formats fit within the original \texttt{4096}-token context window. It contains 10 CWE categories. For each CWE category, we select 10 Java files from Juliet. Each selected file contains both vulnerable and non-vulnerable code variants. During preprocessing, the extraction script separates these variants and stores them as individual samples. This results in 20 cleaned samples per CWE category, consisting of 10 vulnerable samples and 10 non-vulnerable samples. In total, the short-sample dataset contains 200 samples.

The selected CWE categories are shown in Table~\ref{tab:dataset-cwes}. The first 10 CWE categories are used in both datasets, while the shaded rows indicate the additional categories included only in the longer-sample dataset.
 
\begin{table}[t]
\centering
\caption{CWE categories included in the datasets. Shaded rows indicate categories used only in the longer-sample dataset.}
\label{tab:dataset-cwes}
\begin{tabular}{p{0.160\linewidth}p{0.72\linewidth}}
\toprule
\textbf{CWE} & \textbf{Name} \\
\midrule
CWE-78 & OS Command Injection \\
CWE-80 & Cross-site Scripting \\
CWE-129 & Improper Validation of Array Index \\
CWE-190 & Integer Overflow \\
CWE-209 & Information Leak Through an Error Message \\
CWE-369 & Divide By Zero \\
CWE-396 & Declaration of Catch for Generic Exception \\
CWE-400 & Uncontrolled Resource Consumption \\
CWE-476 & NULL Pointer Dereference \\
CWE-470 & Use of Externally-Controlled Input to Select Classes or Code \\

\midrule
\rowcolor{gray!15} CWE-89 & SQL Injection \\
\rowcolor{gray!15} CWE-90 & LDAP Injection \\
\rowcolor{gray!15} CWE-327 & Use of Broken or Risky Cryptographic Algorithm \\

\bottomrule
\end{tabular}
\end{table}

The second dataset is the longer-sample dataset, which we use to evaluate larger code samples where the source code and BT inputs remain processable, while many corresponding AST inputs exceed the original \texttt{4096}-token context limit after conversion. For the 10 main CWE categories, we select a different set of longer Juliet Java samples than the ones used in the short-sample dataset. For each CWE category, the extraction script processes 10 selected files in the same way, resulting in 20 cleaned samples per category and 200 samples across the 10 main categories.

In addition, we include three extended CWE categories: CWE-327, CWE-89, and CWE-90. For each extended CWE category, the extraction script processes 10 selected files in the same way, resulting in 20 cleaned samples per category. Overall, the longer-sample dataset contains 13 CWE categories and 260 cleaned samples. In this setting, we compare only source code in its original form and BT representations, since the corresponding AST representations exceed the context limit after conversion. In this setting, we compare source code in its original form and BT representations as full-coverage inputs, while AST representations are considered separately in the context-window sensitivity experiment to evaluate their input coverage under larger context windows.

\subsection{Evaluation Strategy}
\label{subsec:evaluation-strategy}

The model outputs are evaluated using three matching strategies. The first strategy is exact CWE matching, where a prediction is considered correct only if the output contains the expected CWE identifier. This is the strictest evaluation setting, as it requires the model to provide the correct CWE identifier explicitly.

The second strategy is keyword matching. In this setting, the output is considered correct if it contains either the expected CWE identifier or a vulnerability name derived from the expected CWE category. This captures cases where the model identifies the correct vulnerability type by name, even if the exact CWE identifier is missing or incorrect.

The third strategy is semantic similarity matching. In this setting, the model output is compared with the official CWE description using an embedding-based similarity approach.\footnote{We use the \texttt{sentence-transformers/all-mpnet-base-v2} embedding model: \url{https://huggingface.co/sentence-transformers/all-mpnet-base-v2}.} This captures cases where the model explanation is semantically close to the expected vulnerability category, even if the exact CWE identifier or vulnerability name is missing or incorrect. Semantic similarity is computed only for vulnerable samples, since non-vulnerable samples do not have an expected CWE category to compare against.

For exact CWE matching and keyword matching, we report accuracy, precision, recall, and F1 score. These metrics are computed from true positives, false positives, true negatives, and false negatives. We report the semantic similarity strategy separately using similarity scores between the model output and the official CWE description, where higher scores indicate greater semantic similarity.

\subsection{Implementation Details}

The implementation follows the framework described in Section~\ref{sec:method} and automates the generation of AST and BT representations from source code samples. For each sample, the pipeline produces three input formats used in the experiments, namely source code in its original form, an AST representation, and a BT representation. 

For each Java source code sample, the implemented conversion module first uses the JavaParser library to generate an AST representation. The generated AST is then converted into a BT representation using the custom AST-to-BT converter developed in this work, following the mapping rules summarized in Table~\ref{tab:ast-bt-mapping}.

After generating the three input formats, the pipeline creates one prompt for each representation of each sample. We employ a role-based prompting strategy, in which the model is assigned the role of a security analyst and tasked with detecting vulnerabilities in the provided input. The same prompt structure is used for all three input representations. Figure~\ref{fig:prompt-template} shows the prompt template used in the experiments. In the template, \texttt{<type>} is replaced with the representation type (code, AST, or BT), and \texttt{<content>} is replaced with the corresponding input content for the sample. Both the AST and BT representations are stored as JSON structures before being inserted into the prompts.

\begin{figure}[t]
\centering
\begin{tcolorbox}[
    width=0.95\linewidth,
    colback=lightgreen,
    colframe=promptdarkgreen,
    boxrule=0.8pt,
    arc=2mm,
    left=2mm,
    right=2mm,
    top=1.5mm,
    bottom=1.5mm,
    enhanced,
    drop shadow={black!12!white}
]

\vspace{0.4em}
You are a security analyst. Analyze the following \texttt{<type>} for vulnerabilities. Return the CWE if possible.

\vspace{0.5em}
\texttt{<content>}
\end{tcolorbox}
\caption{Prompt template used in the experiments.}
\label{fig:prompt-template}
\end{figure}

The generated prompts are used as input to the \texttt{Q4\_K\_M} GGUF quantization of Mistral Small 3.2 24B Instruct 2506.\footnote{\url{https://huggingface.co/bartowski/mistralai_Mistral-Small-3.2-24B-Instruct-2506-GGUF}} The model is run with temperature set to 0.0, top-$p$ set to 0.9, and repetition penalty set to 1.15. We set the initial context window to \texttt{4096} tokens and the maximum number of generated tokens to \texttt{1024}. The context window includes both the input prompt and the generated output. The same decoding configuration is used for all input representations to ensure a controlled comparison. The generated model outputs are then evaluated using the strategies described in Subsection~\ref{subsec:evaluation-strategy}.

\section{Experimental Results}
\label{sec:results}

In this section, we present the results for the three experiments conducted in this study. In the first experiment, we evaluate the vulnerability detection performance of the LLM on the short-sample dataset when the input is provided in three different formats: source code in its original form, AST, and BT.

In the second experiment, we evaluate the LLM on the longer-sample dataset using source code in its original form and BT inputs. AST inputs are not included in this comparison because many longer AST representations exceed the original \texttt{4096}-token context window.

In the third experiment, we conduct a context-window sensitivity experiment on the longer-sample dataset. This experiment evaluates how increasing the context window affects the number of processable inputs, especially for longer AST representations.

The experiments use the same model, prompt template, decoding configuration, and evaluation strategies described in Subsection~\ref{subsec:evaluation-strategy}. The short-sample and longer-sample experiments use the original context window of \texttt{4096} tokens and are repeated across 10 runs using the same configuration. The reported values for these two experiments are computed across 10 runs. Across the repeated runs, the standard deviation is 0.00 for all reported metrics, indicating that the repeated executions produced identical evaluation results under the fixed decoding configuration. For this reason, standard deviation values are not included in the result tables. The context-window sensitivity experiment is conducted once for each context-window size on the longer-sample dataset.

\subsection{Short-Sample Representation Comparison}

The short-sample experiment compares source code in its original form, AST, and BT under the same context conditions, since all three representations fit within the context limit for this dataset. Table~\ref{tab:short-overall-results} summarizes the overall results. Under both matching strategies, BT achieves the highest recall, indicating that it identifies the largest number of vulnerable samples.
However, source code in its original form achieves the highest precision and accuracy under both matching strategies. This suggests that BT improves detection coverage, while source code in its original form remains more precise in this setting. Overall, the results show a trade-off between recall and precision. The AST representation does not show a consistent advantage over the other representations in this setting.

\begin{table}[t]
    \centering
    \caption{Overall results by input representation on the short-sample dataset. Semantic similarity is reported as mean similarity.}
    \label{tab:short-overall-results}
    \renewcommand{\arraystretch}{1.1}
    \small
    \begin{tabular}{lcccc}
    \toprule
    \rowcolor{gray!15}\textbf{Input repr.} & \textbf{Precision} & \textbf{Recall} & \textbf{F1} & \textbf{Accuracy} \\
    \midrule
    \multicolumn{5}{l}{\textbf{Keyword matching}} \\
    Source Code & 0.60 & 0.75 & 0.67 & 0.63 \\
    AST & 0.55 & 0.72 & 0.62 & 0.56 \\
    BT & 0.57 & \cellcolor{green!15}0.83 & 0.67 & 0.60 \\
    \midrule
    \multicolumn{5}{l}{\textbf{Exact CWE matching}} \\
    Source Code & 0.59 & 0.54 & 0.57 & 0.59 \\
    AST & 0.55 & 0.59 & 0.57 & 0.55 \\
    BT & 0.56 & \cellcolor{green!15}0.68 & 0.62 & 0.58 \\
  
    \bottomrule
    \end{tabular}
\end{table}

% The overall mean semantic similarity scores are also close across the three input representations, with source code in its original form slightly higher than AST and BT in this setting. 
 % \multicolumn{5}{l}{\textbf{Semantic similarity}} \\
 %    Source Code & \multicolumn{4}{c}{\colorbox{green!15}{0.4660}} \\
 %    AST & \multicolumn{4}{c}{0.4652} \\
 %    BT & \multicolumn{4}{c}{0.4560} \\

Examining the individual CWE categories offers a more detailed view of these trends. Table~\ref{tab:short-per-cwe-results} reports the per-CWE results for both matching strategies. The best-performing representation varies across CWE categories, showing that no single representation is consistently strongest for all weakness types. BT achieves the highest or tied-highest recall for several CWE categories and also obtains the highest F1 score in several cases. Source code in its original form performs best for other categories, while AST shows stronger results only in a few cases. Overall, the per-CWE results indicate that the relative performance of each representation varies by CWE category.

\begin{table*}[t]
\centering
\caption{Per-CWE results for keyword and exact CWE matching on the short-sample dataset. Shaded F1 values indicate the highest F1 score for each CWE and matching strategy.}
\label{tab:short-per-cwe-results}

\setlength{\tabcolsep}{12pt}
\renewcommand{\arraystretch}{0.7}
\setlength{\aboverulesep}{0.3ex}
\setlength{\belowrulesep}{0.3ex}

\begin{tabular}{llcccccccc}
\toprule
\multirow{2}{*}{\textbf{CWE}}
& \multirow{2}{*}{\textbf{Input Repr.}}
& \multicolumn{4}{c}{\textbf{Keyword matching}}
& \multicolumn{4}{c}{\textbf{Exact CWE matching}} \\
\cmidrule(lr){3-6}
\cmidrule(lr){7-10}
& & \textbf{Prec.} & \textbf{Rec.} & \textbf{F1} & \textbf{Acc.}
  & \textbf{Prec.} & \textbf{Rec.} & \textbf{F1} & \textbf{Acc.} \\
\midrule

\multirow{3}{*}{CWE-78}
& Source Code & 0.56 & 1.00 & 0.71 & 0.60
              & 0.56 & 1.00 & 0.71 & 0.60 \\
& AST         & 0.53 & 1.00 & 0.69 & 0.55
              & 0.53 & 1.00 & 0.69 & 0.55 \\
& BT          & 0.69 & 0.90 & \cellcolor{green!15}0.78 & 0.75
              & 0.69 & 0.90 & \cellcolor{green!15}0.78 & 0.75 \\
\cmidrule(lr){1-10}

\multirow{3}{*}{CWE-80}
& Source Code & 0.56 & 0.90 & 0.69 & 0.60
              & 1.00 & 0.20 & \cellcolor{green!15}0.33 & 0.60 \\
& AST         & 0.59 & 1.00 & \cellcolor{green!15}0.74 & 0.65
              & 0.00 & 0.00 & 0.00 & 0.50 \\
& BT          & 0.56 & 1.00 & 0.71 & 0.60
              & 0.00 & 0.00 & 0.00 & 0.40 \\
\cmidrule(lr){1-10}

\multirow{3}{*}{CWE-129}
& Source Code & 0.53 & 0.80 & 0.64 & 0.55
              & 0.50 & 0.40 & 0.44 & 0.50 \\
& AST         & 0.57 & 0.80 & 0.67 & 0.60
              & 0.62 & 0.80 & \cellcolor{green!15}0.70 & 0.65 \\
& BT          & 0.59 & 1.00 & \cellcolor{green!15}0.74 & 0.65
              & 0.53 & 0.80 & 0.64 & 0.55 \\
\cmidrule(lr){1-10}

\multirow{3}{*}{CWE-190}
& Source Code & 0.62 & 1.00 & \cellcolor{green!15}0.77 & 0.70
              & 0.58 & 0.70 & \cellcolor{green!15}0.64 & 0.60 \\
& AST         & 0.46 & 0.60 & 0.52 & 0.45
              & 0.50 & 0.60 & 0.55 & 0.50 \\
& BT          & 0.50 & 1.00 & 0.67 & 0.50
              & 0.50 & 0.90 & \cellcolor{green!15}0.64 & 0.50 \\
\cmidrule(lr){1-10}

\multirow{3}{*}{CWE-209}
& Source Code & 1.00 & 0.70 & \cellcolor{green!15}0.82 & 0.85
              & 1.00 & 0.20 & 0.33 & 0.60 \\
& AST         & 0.50 & 0.30 & 0.38 & 0.50
              & 0.67 & 0.20 & 0.31 & 0.55 \\
& BT          & 0.62 & 1.00 & 0.77 & 0.70
              & 0.75 & 0.90 & \cellcolor{green!15}0.82 & 0.80 \\
\cmidrule(lr){1-10}

\multirow{3}{*}{CWE-369}
& Source Code & 0.53 & 0.80 & \cellcolor{green!15}0.64 & 0.55
              & 0.57 & 0.80 & \cellcolor{green!15}0.67 & 0.60 \\
& AST         & 0.47 & 0.70 & 0.56 & 0.45
              & 0.47 & 0.70 & 0.56 & 0.45 \\
& BT          & 0.42 & 0.50 & 0.45 & 0.40
              & 0.42 & 0.50 & 0.45 & 0.40 \\
\cmidrule(lr){1-10}

\multirow{3}{*}{CWE-396}
& Source Code & 1.00 & 0.10 & 0.18 & 0.55
              & 1.00 & 0.10 & 0.18 & 0.55 \\
& AST         & 0.80 & 0.80 & \cellcolor{green!15}0.80 & 0.80
              & 0.80 & 0.80 & \cellcolor{green!15}0.80 & 0.80 \\
& BT          & 0.83 & 0.50 & 0.62 & 0.70
              & 0.83 & 0.50 & 0.62 & 0.70 \\
\cmidrule(lr){1-10}

\multirow{3}{*}{CWE-400}
& Source Code & 0.67 & 0.80 & \cellcolor{green!15}0.73 & 0.70
              & 0.67 & 0.60 & \cellcolor{green!15}0.63 & 0.65 \\
& AST         & 0.50 & 0.70 & 0.58 & 0.50
              & 0.46 & 0.60 & 0.52 & 0.45 \\
& BT          & 0.50 & 0.80 & 0.62 & 0.50
              & 0.50 & 0.70 & 0.58 & 0.50 \\
\cmidrule(lr){1-10}

\multirow{3}{*}{CWE-470}
& Source Code & 0.44 & 0.40 & 0.42 & 0.45
              & 0.44 & 0.40 & 0.42 & 0.45 \\
& AST         & 0.45 & 0.50 & 0.48 & 0.45
              & 0.45 & 0.50 & 0.48 & 0.45 \\
& BT          & 0.50 & 0.70 & \cellcolor{green!15}0.58 & 0.50
              & 0.50 & 0.70 & \cellcolor{green!15}0.58 & 0.50 \\
\cmidrule(lr){1-10}

\multirow{3}{*}{CWE-476}
& Source Code & 0.62 & 1.00 & \cellcolor{green!15}0.77 & 0.70
              & 0.62 & 1.00 & \cellcolor{green!15}0.77 & 0.70 \\
& AST         & 0.62 & 0.80 & 0.70 & 0.65
              & 0.58 & 0.70 & 0.64 & 0.60 \\
& BT          & 0.60 & 0.90 & 0.72 & 0.65
              & 0.60 & 0.90 & 0.72 & 0.65 \\

\bottomrule
\end{tabular}
\end{table*}

In addition to exact CWE and keyword matching, we also evaluate the semantic similarity of the generated outputs to the expected CWE descriptions.
Table~\ref{tab:semantic-per-cwe} reports the mean semantic similarity scores for each CWE category and input representation. Similar to the previous results, the representation with the highest mean semantic similarity varies across CWE categories.
Overall, the mean semantic similarity scores are very close across the three input representations, with source code in its original form obtaining the highest mean score of 0.4660, followed by AST with 0.4652 and BT with 0.4560. The standard deviation values are also similar, indicating comparable variation across CWE categories for all three representations.

\begin{table}[t]
\centering
\caption{Per-CWE mean semantic similarity scores on the short-sample dataset. Shaded cells indicate the highest score for each CWE.}
\label{tab:semantic-per-cwe}

\setlength{\tabcolsep}{8pt}
\renewcommand{\arraystretch}{1.1}
\begin{tabular}{lccc}
\toprule
\rowcolor{gray!15} \textbf{CWE} & \textbf{Source Code} & \textbf{AST} & \textbf{BT} \\
\midrule
CWE-78  & 0.3795 & 0.4046 & \cellcolor{green!15}0.4472 \\
CWE-80  & \cellcolor{green!15}0.6061 & 0.5808 & 0.5985 \\
CWE-129 & \cellcolor{green!15}0.6409 & 0.6310 & 0.6100 \\
CWE-190 & 0.4332 & 0.4139 & \cellcolor{green!15}0.4499 \\
CWE-209 & 0.2712 & \cellcolor{green!15}0.2794 & 0.2722 \\
CWE-369 & \cellcolor{green!15}0.2748 & 0.2214 & 0.2103 \\
CWE-396 & 0.5306 & \cellcolor{green!15}0.5911 & 0.5131 \\
CWE-400 & 0.4653 & \cellcolor{green!15}0.4871 & 0.4636 \\
CWE-470 & \cellcolor{green!15}0.4668 & 0.4462 & 0.4376 \\
CWE-476 & 0.5913 & \cellcolor{green!15}0.5964 & 0.5578 \\

\midrule
Mean & \cellcolor{green!15}0.4660 & 0.4652 & 0.4560 \\
Std. dev. & 0.1305 & 0.1395 & 0.1301 \\

\bottomrule
\end{tabular}
\end{table}

\subsection{Longer-Sample Code and BT Comparison}

The longer-sample experiment compares source code in its original form and BT representations, since the AST representations exceed the context limit. Table~\ref{tab:long-overall-results} summarizes the overall results. Compared with source code in its original form, BT achieves higher precision, recall, F1 score, and accuracy under both keyword matching and exact CWE matching. The largest difference is observed in recall, where BT increases from 0.73 to 0.83 under keyword matching and from 0.63 to 0.69 under exact CWE matching. 

\begin{table}[t]
    \centering
    \caption{Overall results by input format on the longer-sample dataset. Semantic similarity is reported as mean similarity.}
    \label{tab:long-overall-results}
    \small
    \renewcommand{\arraystretch}{1.1}
    \begin{tabular}{lcccc}
    \toprule
    \rowcolor{gray!15} \textbf{Input format} & \textbf{Precision} & \textbf{Recall} & \textbf{F1} & \textbf{Accuracy} \\
    \midrule
    \multicolumn{5}{l}{\textbf{Keyword matching}} \\
    Source Code & 0.49 & 0.73 & 0.59 & 0.49 \\
    BT & \cellcolor{green!15}0.51 & \cellcolor{green!15}0.83 & \cellcolor{green!15}0.64 & \cellcolor{green!15}0.52 \\
    \midrule
    \multicolumn{5}{l}{\textbf{Exact CWE matching}} \\
    Source Code & 0.50 & 0.63 & 0.56 & 0.50 \\
    BT & \cellcolor{green!15}0.51 & \cellcolor{green!15}0.69 & \cellcolor{green!15}0.58 & \cellcolor{green!15}0.51 \\
 
    \bottomrule
    \end{tabular}
\end{table}

% The semantic similarity results show the same overall direction, with BT obtaining a slightly higher mean similarity score than the source code in its original form.
 % \midrule
 %    \multicolumn{5}{l}{\textbf{Semantic similarity}} \\
 %    Source Code & \multicolumn{4}{c}{0.4411} \\
 %    BT & \multicolumn{4}{c}{\cellcolor{green!15}0.4470} \\ 

The per-CWE results in Table~\ref{tab:long-per-cwe-results} provide a more detailed view of how the two input representations compare across different weakness types. BT achieves the highest F1 score for eight categories, while the source code in its original form performs better for the other five categories. Overall, the longer-sample results suggest that BT can improve detection performance for longer inputs, with the strongest gains observed for specific CWE categories.

\begin{table*}[t]
\centering
\caption{Per-CWE results for keyword and exact CWE matching on the longer-sample dataset. Shaded F1 values indicate the highest F1 score for each CWE and matching strategy.}
\label{tab:long-per-cwe-results}

\setlength{\tabcolsep}{12pt}
\renewcommand{\arraystretch}{0.7}
\setlength{\aboverulesep}{0.3ex}
\setlength{\belowrulesep}{0.3ex}

\begin{tabular}{llcccccccc}
\toprule
\multirow{2}{*}{\textbf{CWE}}
& \multirow{2}{*}{\textbf{Input Repr.}}
& \multicolumn{4}{c}{\textbf{Keyword matching}}
& \multicolumn{4}{c}{\textbf{Exact CWE matching}} \\
\cmidrule(lr){3-6}
\cmidrule(lr){7-10}
& & \textbf{Prec.} & \textbf{Rec.} & \textbf{F1} & \textbf{Acc.}
  & \textbf{Prec.} & \textbf{Rec.} & \textbf{F1} & \textbf{Acc.} \\
\midrule

\multirow{2}{*}{CWE-78}
& Source Code & 0.50 & 1.00 & 0.67 & 0.50
              & 0.50 & 1.00 & 0.67 & 0.50 \\
& BT          & 0.59 & 1.00 & \cellcolor{green!15}0.74 & 0.65
              & 0.59 & 1.00 & \cellcolor{green!15}0.74 & 0.65 \\
\cmidrule(lr){1-10}

\multirow{2}{*}{CWE-80}
& Source Code & 0.44 & 0.80 & 0.57 & 0.40
              & 0.00 & 0.00 & 0.00 & 0.50 \\
& BT          & 0.50 & 0.90 & \cellcolor{green!15}0.64 & 0.50
              & 0.50 & 0.10 & \cellcolor{green!15}0.18 & 0.55 \\
\cmidrule(lr){1-10}

\multirow{2}{*}{CWE-89}
& Source Code & 0.53 & 1.00 & \cellcolor{green!15}0.69 & 0.55
              & 0.53 & 1.00 & \cellcolor{green!15}0.69 & 0.55 \\
& BT          & 0.50 & 1.00 & 0.67 & 0.50
              & 0.50 & 1.00 & 0.67 & 0.50 \\
\cmidrule(lr){1-10}

\multirow{2}{*}{CWE-90}
& Source Code & 0.53 & 0.90 & \cellcolor{green!15}0.67 & 0.55
              & 0.53 & 0.90 & \cellcolor{green!15}0.67 & 0.55 \\
& BT          & 0.50 & 0.90 & 0.64 & 0.50
              & 0.35 & 0.30 & 0.32 & 0.35 \\
\cmidrule(lr){1-10}

\multirow{2}{*}{CWE-129}
& Source Code & 0.47 & 0.70 & 0.56 & 0.45
              & 0.40 & 0.40 & 0.40 & 0.40 \\
& BT          & 0.53 & 0.80 & \cellcolor{green!15}0.64 & 0.55
              & 0.50 & 0.70 & \cellcolor{green!15}0.58 & 0.50 \\
\cmidrule(lr){1-10}

\multirow{2}{*}{CWE-190}
& Source Code & 0.53 & 1.00 & \cellcolor{green!15}0.69 & 0.55
              & 0.53 & 1.00 & \cellcolor{green!15}0.69 & 0.55 \\
& BT          & 0.47 & 0.90 & 0.62 & 0.45
              & 0.47 & 0.90 & 0.62 & 0.45 \\
\cmidrule(lr){1-10}

\multirow{2}{*}{CWE-209}
& Source Code & 0.70 & 0.70 & \cellcolor{green!15}0.70 & 0.70
              & 0.75 & 0.60 & \cellcolor{green!15}0.67 & 0.70 \\
& BT          & 0.53 & 0.90 & 0.67 & 0.55
              & 0.53 & 0.90 & \cellcolor{green!15}0.67 & 0.55 \\
\cmidrule(lr){1-10}

\multirow{2}{*}{CWE-327}
& Source Code & 0.36 & 0.50 & 0.42 & 0.30
              & 0.36 & 0.50 & 0.42 & 0.30 \\
& BT          & 0.53 & 1.00 & \cellcolor{green!15}0.69 & 0.55
              & 0.53 & 1.00 & \cellcolor{green!15}0.69 & 0.55 \\
\cmidrule(lr){1-10}

\multirow{2}{*}{CWE-369}
& Source Code & 0.40 & 0.60 & 0.48 & 0.35
              & 0.40 & 0.60 & 0.48 & 0.35 \\
& BT          & 0.59 & 1.00 & \cellcolor{green!15}0.74 & 0.65
              & 0.59 & 1.00 & \cellcolor{green!15}0.74 & 0.65 \\
\cmidrule(lr){1-10}

\multirow{2}{*}{CWE-396}
& Source Code & 1.00 & 0.70 & \cellcolor{green!15}0.82 & 0.85
              & 1.00 & 0.70 & \cellcolor{green!15}0.82 & 0.85 \\
& BT          & 0.71 & 0.50 & 0.59 & 0.65
              & 0.71 & 0.50 & 0.59 & 0.65 \\
\cmidrule(lr){1-10}

\multirow{2}{*}{CWE-400}
& Source Code & 0.53 & 0.80 & \cellcolor{green!15}0.64 & 0.55
              & 0.58 & 0.70 & \cellcolor{green!15}0.64 & 0.60 \\
& BT          & 0.42 & 0.50 & 0.45 & 0.40
              & 0.30 & 0.20 & 0.24 & 0.35 \\
\cmidrule(lr){1-10}

\multirow{2}{*}{CWE-470}
& Source Code & 0.17 & 0.10 & 0.12 & 0.30
              & 0.20 & 0.10 & 0.13 & 0.35 \\
& BT          & 0.43 & 0.60 & \cellcolor{green!15}0.50 & 0.40
              & 0.43 & 0.60 & \cellcolor{green!15}0.50 & 0.40 \\
\cmidrule(lr){1-10}

\multirow{2}{*}{CWE-476}
& Source Code & 0.41 & 0.70 & 0.52 & 0.35
              & 0.41 & 0.70 & 0.52 & 0.35 \\
& BT          & 0.47 & 0.80 & \cellcolor{green!15}0.59 & 0.45
              & 0.47 & 0.80 & \cellcolor{green!15}0.59 & 0.45 \\

\bottomrule
\end{tabular}
\end{table*}

The semantic similarity results provide a similar comparison from the perspective of output meaning. Table~\ref{tab:long-semantic-per-cwe} reports the per-CWE mean scores for each input representation. BT achieves a higher mean similarity score for 8 of the 13 CWE categories, while the source code in its original form achieves a higher score for the remaining 5 categories. These results show that BT remains competitive for longer samples, with variation across CWE categories. Overall, BT obtains a slightly higher mean semantic similarity score than source code in its original form, with 0.4470 compared with 0.4411. BT also has a lower standard deviation, 0.1503, compared with 0.1694, suggesting slightly less variation across CWE categories.

\begin{table}[t]
\centering
\caption{Per-CWE mean semantic similarity scores on the longer-sample dataset. Shaded cells indicate the highest score for each CWE.}
\label{tab:long-semantic-per-cwe}
\setlength{\tabcolsep}{8pt}
\renewcommand{\arraystretch}{1.1}
\begin{tabular}{lcc}
\toprule
\rowcolor{gray!15} \textbf{CWE} & \textbf{Source Code} & \textbf{BT} \\
\midrule
CWE-78  & 0.3473 & \cellcolor{green!15}0.3842 \\
CWE-80  & \cellcolor{green!15}0.5351 & 0.4641 \\
CWE-89  & \cellcolor{green!15}0.6747 & 0.6642 \\
CWE-90  & \cellcolor{green!15}0.7259 & 0.7166 \\
CWE-129 & \cellcolor{green!15}0.5998 & 0.5678 \\
CWE-190 & 0.2908 & \cellcolor{green!15}0.3615 \\
CWE-209 & 0.2922 & \cellcolor{green!15}0.3068 \\
CWE-327 & 0.2176 & \cellcolor{green!15}0.2520 \\
CWE-369 & 0.2045 & \cellcolor{green!15}0.2162 \\
CWE-396 & 0.4974 & \cellcolor{green!15}0.5237 \\
CWE-400 & 0.4308 & \cellcolor{green!15}0.4447 \\
CWE-470 & 0.3845 & \cellcolor{green!15}0.3910 \\
CWE-476 & \cellcolor{green!15}0.5342 & 0.5178 \\
\midrule
Mean & 0.4411 & \cellcolor{green!15}0.4470 \\
Std. dev. & 0.1694 & 0.1503 \\
\bottomrule
\end{tabular}
\end{table}

\subsection{Context Window Sensitivity}

The context-window sensitivity experiment evaluates whether increasing the context window improves the practicality of using longer input representations. This experiment focuses primarily on representation feasibility, since longer AST representations can exceed the available context limit after conversion. We compare the original context window of \texttt{4096} tokens with larger context windows of \texttt{8192} and \texttt{16384} tokens.

Table~\ref{tab:context-coverage} reports the number of processable inputs for each representation across the three context-window sizes. Source code in its original form and BT representations remain fully processable across all context-window sizes, with all 260 inputs fitting within the context limit. For AST representations, coverage increases as the context window becomes larger, from 86 out of 260 inputs at \texttt{4096} tokens to 156 inputs at \texttt{8192} tokens and 222 inputs at \texttt{16384} tokens. However, even with the largest context window tested, not all AST inputs fit within the available context.

These results show that increasing the context window improves the feasibility of using AST representations, but does not fully solve the input-length problem for the longer-sample dataset. BT representations remain more practical in this setting, since they preserve full input coverage even at the original context-window size.

\begin{table}[t]
\centering
\caption{Input coverage across context-window sizes on the longer-sample dataset.}
\label{tab:context-coverage}
\setlength{\tabcolsep}{8pt}
\renewcommand{\arraystretch}{1.2}
\begin{tabular}{lccc}
\toprule
\rowcolor{gray!15} \textbf{Representation} & \textbf{4096} & \textbf{8192} & \textbf{16384} \\
\midrule
Source Code & 260/260 & 260/260 & 260/260 \\ 
AST & 86/260 & 156/260 & 222/260   \\ 
BT & 260/260 & 260/260 & 260/260   \\ 
\bottomrule
\end{tabular}
\end{table}

The evaluation metrics also show small variations when the context window is increased. For example, under exact CWE matching, the overall F1-score for BT is 0.58 at the original \texttt{4096}-token setting and remains almost the same at \texttt{16384} tokens, with an F1-score of 0.58. However, it decreases to 0.54 at \texttt{8192} tokens. Similarly, the F1-score for source code in its original form is 0.56 at both \texttt{4096} and \texttt{16384} tokens, and increases slightly to 0.58 at \texttt{8192} tokens. These results suggest that increasing the context window mainly affects input coverage, especially for AST representations, while the overall detection performance for source code and BT remains relatively stable.

\section{Discussion}
\label{sec:discussion}

In this section, we discuss the main findings from the three experimental settings. First, we examine how the choice of input representation affects vulnerability detection when all representations fit within the context limit. We then discuss the longer-sample setting, where BT is compared with source code in its original form and AST representations are limited by input length. Finally, we discuss the effect of increasing the context window on the practicality of structured representations. 
The discussion refers back to the corresponding result subsections and tables in Section~\ref{sec:results}.

\subsection{Effect of Input Representation on Short Samples}

The short-sample results in Table~\ref{tab:short-overall-results} show that the choice of input representation affects LLM-based vulnerability detection. In this setting, all three representations fit within the context limit, so the comparison focuses on the effect of representation format rather than input-length limitations. 
BT achieves the highest recall in the overall short-sample results, which suggests that the behavior-oriented structure helps the model identify more vulnerable samples. At the same time, source code in its original form achieves higher precision and accuracy, showing that BT also produces more false positives in this setting. This indicates a trade-off between detection coverage and precision.

The AST representation does not show a consistent advantage in the short-sample experiment. Although ASTs provide detailed syntactic information, this detail does not necessarily improve vulnerability detection for the lightweight LLM used in this study.

The per-CWE results in Table~\ref{tab:short-per-cwe-results} suggest that BT may be especially useful for weaknesses where the relevant behavior is reflected in conditions, actions, or the ordering of operations. For example, BT achieves strong results for CWE-129, CWE-470, CWE-209, and CWE-78, where vulnerability detection can depend on how inputs are checked, propagated, or used in executable operations. However, this pattern should be interpreted cautiously, since the results also vary across categories and the sample size per CWE is limited.

This variation also shows that BT is not uniformly better across all CWE categories. Source code in its original form performs better for some weaknesses, and AST shows stronger results in selected cases. Therefore, the results should not be interpreted as showing that BT is always superior. Instead, they suggest that BT can be a useful alternative representation, especially when behavior-oriented structure helps expose vulnerability-relevant information. Overall, the short-sample experiment suggests that BT is most useful as a recall-oriented representation rather than as a uniformly better replacement for source code in its original form or AST.

\subsection{Behavior Trees for Longer Samples}  
The longer-sample results in Table~\ref{tab:long-overall-results} provide a clearer view of the practical value of BT representations. Unlike the short-sample setting, the AST representation cannot be used consistently for longer samples because many AST inputs exceed the context limit. This shows a practical limitation of using AST representations directly as LLM inputs.

In this setting, BT remains processable for all longer samples and achieves better overall results than source code in its original form under both keyword matching and exact CWE matching. The overall results show that BT improves the F1-score from 0.59 to 0.64 under keyword matching and from 0.56 to 0.58 under exact CWE matching. BT also achieves higher recall in both settings, increasing from 0.73 to 0.83 under keyword matching and from 0.63 to 0.69 under exact CWE matching. These results suggest that BT improves detection coverage on the longer-sample dataset, with the clearest difference observed in recall.

The per-CWE results in Table~\ref{tab:long-per-cwe-results} show that the benefit of BT is not limited to one weakness category. BT achieves stronger F1-scores for several categories, including CWE-78, CWE-80, CWE-129, CWE-327, CWE-369, CWE-470, and CWE-476. The semantic similarity results in Table~\ref{tab:long-semantic-per-cwe} also support this trend, with BT achieving higher mean similarity scores for 8 of the 13 CWE categories. However, the absolute semantic similarity scores should be interpreted cautiously, since the model outputs are practical vulnerability analyses, while the CWE descriptions are formal category-level definitions. As a result, the scores are more useful for comparing representations under the same evaluation setting than for judging output quality in isolation. These results suggest that BT can provide a useful intermediate representation for longer samples, especially when detailed AST representations become too verbose.
  
\subsection{Context-Window Feasibility}

The context-window results in Table~\ref{tab:context-coverage} show that increasing the context window improves the feasibility of using longer AST representations, but does not remove the practicality problem entirely. When the context window is increased from \texttt{4096} to \texttt{8192} and \texttt{16384} tokens, more AST inputs become processable. However, even at \texttt{16384} tokens, the AST representation still does not cover the full longer-sample dataset. This shows that simply increasing the context window is not enough to make verbose structured representations consistently practical.

This finding is notable because ASTs preserve detailed syntactic structure, but this detail comes with a large input-size cost. For longer samples, the AST representation can become too large to fit into the model context, even when the source code and BT representations remain fully processable. In contrast, BT provides a more compact structured representation that remains usable across all context-window settings tested in this work.

From a practical perspective, increasing the context window also requires more computational resources and processing time. This makes larger-context experiments more expensive, especially when running local lightweight models across many samples and repeated runs. Therefore, relying only on larger context windows is not always an efficient solution. BT provides an alternative by reducing the input size while still preserving a behavior-oriented structure that is useful for vulnerability detection.

Overall, the context-window results strengthen the motivation for using BT as an intermediate representation. Compared with AST, BT is more compact and remains fully processable even under smaller context limits.

\section{Threats to Validity}
\label{sec:threats}

This section discusses the main threats to validity of this study. 
First, the evaluation is based on synthetic Java samples from the Juliet test suite. Although Juliet provides controlled examples with known CWE labels, the samples may not fully reflect the complexity of real-world software projects.

Another limitation concerns the dataset size and selection. The experiments use selected CWE categories and balanced vulnerable and non-vulnerable samples. This makes the comparison between input representations easier to control, but it also limits how broadly the results can be generalized.

The evaluation is also limited to one lightweight LLM and one prompt template. Since LLM-based vulnerability detection can be sensitive to the model, prompt wording, and decoding configuration, the results may differ with other models or prompting strategies. We reduce this threat by keeping the model configuration and prompt template fixed across all input representations, which allows the comparison to focus on the effect of the input format rather than changes in the model or prompt setup.

Finally, the evaluation metrics have some limitations. Exact CWE matching can be strict because CWE categories may have parent-child relationships. For example, predicting CWE-79 instead of its child CWE-80 may identify the broader XSS weakness but still be counted as incorrect.
Therefore, a model output may be partially correct from a security perspective while still being counted as incorrect under exact CWE matching. To reduce this threat, we also include keyword matching and semantic similarity matching. However, these metrics also have limitations, keyword matching depends on the official CWE name, while semantic similarity depends on the embedding model and the official CWE descriptions used for comparison. Semantic similarity is also computed only for vulnerable samples, since non-vulnerable samples do not have an expected CWE category.

\section{Conclusion and Future Work}
\label{sec:conclusion}
 
This paper presented a preliminary investigation of BTs as an intermediate representation for LLM-based vulnerability detection. We proposed a BT-based preprocessing framework that transforms Java source code into behavior-oriented tree representations and evaluated this representation against source code in its original form and AST representations using a lightweight LLM.

The results show that BT representations can be useful for vulnerability detection, especially for longer samples. On the short-sample dataset, BT achieved the highest recall, although source code in its original form achieved higher precision and accuracy. This indicates a trade-off between detection coverage and precision. On the longer-sample dataset, BT achieved stronger overall results than source code in its original form under both keyword matching and exact CWE matching, while also remaining fully processable within the context window. In contrast, AST representations became less practical for longer samples because many inputs exceeded the available context limit. 

The context-window sensitivity experiment further showed that increasing the context window improves AST coverage, but does not fully solve the input-length problem. BT representations remained fully processable across all tested context-window sizes, suggesting that they provide a more compact structured representation for this setting.

Future work should evaluate the proposed approach on larger and more realistic datasets, including real-world software projects and additional CWE categories. It would also be useful to study BT representations across other programming languages, models, and prompt designs. Finally, future studies could investigate hybrid prompting strategies that combine source code in its original form with BT representations to examine whether this can balance precision and recall.

% \section*{Declaration on the Use of Generative AI}

% Generative AI tools, including LLMs, were used during the preparation of this paper. They were used to assist with the development and debugging of experimental source code, the construction and refinement of scripts used in the experiments, and the polishing of the manuscript text for clarity, grammar, and structure. The research design, experimental decisions, source code used in the experiments, analysis of results, and final content of the paper remain the responsibility of the authors.

\begin{acks}

This work has been supported by the Industrial Graduate School Svensk företagsforskarskola i Cybersäkerhet (SIGS-CyberSec) funded by the Swedish Knowledge Foundation Dnr:20220129; the Wallenberg AI, Autonomous Systems and Software Program (WASP) funded by the Knut and Alice Wallenberg Foundation; and Epiroc Rock Drills AB. We would also like to express our gratitude to Markus Kuchler for his continuous support throughout the course of our research. His participation has been invaluable to the progress of this work.

\end{acks}

%%
%% The next two lines define the bibliography style to be used, and
%% the bibliography file.

\bibliographystyle{ACM-Reference-Format}
\bibliography{references}

\end{document}